\documentstyle[prl,aps,floats,twocolumn,graphics,epsfig]{revtex}

\title{Cluster Growth in a Growing Tree Network}

\author{David Lancaster}

\address{
School of Computer Science, University of Westminster,
Harrow, HA1 3TP. UK.}

\date{\today}
\draft

\begin{document}
\twocolumn[ \hsize\textwidth\columnwidth\hsize\csname
@twocolumnfalse\endcsname 
\maketitle

\begin{abstract}
We analyze a simple model for growing tree networks and find
that although it never percolates, there is an anomalously
large cluster at finite size. We study the growth of
both the maximal cluster and the cluster containing the original
vertex and find that they obey power laws. This
property is also observed through simulations in a non-linear model 
with loops and a true percolating phase.
\end{abstract}

\vspace{1cm}

]

\section{Introduction}
\label{Intro}

Recently there has been interest in the properties of random networks
that are constructed by a growing process. These networks appear to
model certain observed systems rather better than the random graphs of
Erd\H{o}s and R\'enyi~\cite{RandomGraph}. Already, two reviews are
available~\cite{Barabasi1,Mendes1}.

At first, interest 
concentrated on the degree distribution. It has been noticed that 
networks such as the world wide web, the Internet backbone
and scientific collaboration graphs have
(at least in some range) a power law degree distribution. 
This is in contrast to the Poisson distribution found in random
graphs. Barab\'asi and Albert~\cite{Barabasi2,Barabasi3} 
noticed that a power law distribution could
be obtained in a grown network with preferential attachment.

The grown nature of the network creates correlations that affect
more than the degree distribution. 
More recently, these other aspects of grown networks 
have been studied:
in particular, the phenomenon
of percolation, that was of great interest in the study
of random graphs.
Some grown models are devised to be fully connected and percolation
can not be studied, but in a recent paper, 
Calloway {\it et al.}~\cite{Newman01a} 
studied percolation in a very simple network growth model. 
Their model introduces a new vertex at every time step, and also,
with probability $\delta$, makes a link between two existing vertices, chosen
at random. Percolation in this model displays some interesting
features that distinguish it from percolation in a corresponding 
random graph with the same degree distribution. For example, the location and
order of the phase transition are modified by the correlations 
present in the grown model. A similar treatment of other models
has been performed by Dorogovtsev{\it et al.}~\cite{Mendes2},
who note that below the percolation transition, the cluster
size distribution has a power law dependence in contrast to the
exponential dependence typical in non-grown models.

In this paper we pursue the investigation of percolation in
grown networks concentrating on models in which
the new vertex introduced at each time step, is itself the
endpoint of the link possibly created in that time step.
Models of this type were in fact the original kind proposed
by Barab\'asi and Albert~\cite{Barabasi2}.
Only later did 
Dorogovtsev and Mendes~\cite{Mendes3}
introduce the other type of model in which vertex and link creation
are decoupled, and which often happens to be more
convenient for calculation. In general there are families
of such models in which more than one link is added per time step.
When two or more links are added
there seems to be little to distinguish the statistical
features of the two families of models and they can be used interchangeably.
However, for the particular case of single link addition, the
models do have a different character because in the case
where the vertex is attached, only tree networks can be
created. For this reason we call this model, that forms the basis
of study in this paper, the ``tree growth model''.

Although most of the physical networks motivating the surge of interest 
in this subject are
not treelike, other examples, such as food webs, would appear 
not to contain, or at least to have a low probability of containing
loops.
However, a significant reason for studying a tree growth
network comes from past experience:
tree graphs have provided a fruitful field for investigating 
percolation in non-grown networks. They have been studied
both in the Physics literature and through the mathematical
field of branching processes. Tree models provide an
infinite dimensional or mean field model that is often tractable 
in a way that finite dimensional models are not.
We shall find that the tree growth model that forms the
basis of this paper is indeed a simple tractable model that
illuminates more complicated scenarios.

The most interesting feature that we shall use this model to expose is the power
law growth of cluster size as the network size increases.
In numerical work (and for the size of many practical networks) 
this feature, and the presence of what appear to
be anomalously large clusters, mask
the lack of strict percolation in the tree growth model.
This is because in random graphs, 
cluster growth below the percolation 
threshold is only logarithmic. 
However, as was pointed out by Dorogovtsev{\it et al.}~\cite{Mendes2},
the power law growth based on the underlying
power law size distribution, makes the whole phase have
scaling  characteristics typical of critical behavior.

We study the properties of clusters in depth and
besides investigating
the distribution of sizes of clusters chosen at random,
we also study the size of the cluster containing the initial point.
This illuminates the 
intuition that there is a highly connected ``old core'' 
that forms the nucleus of the large clusters and turns out to 
give a useful analytic handle that is not so obvious in non-linear
models with loops. To ensure that the phenomena we are studying are
not an artifact of the tree model we introduce a non-linear extension
and perform some numerical simulations.

It is useful to contrast the properties of this grown tree
model with a non-grown or static analog.
In the present case, we argue
that the appropriate analog
is a branching process rather than a random graph.
Calloway {\it et al.} in their
paper on percolation~\cite{Newman01a}, 
ascribe the cause of the differences between percolation on
grown networks and random graphs to correlations between
the degrees of vertices at each end of connecting links. 
We demonstrate that the tree growth model does not have any such correlations.

The paper is organized as follows. After defining the tree
growth model, the branching process we use as a static analog
is introduced. The percolative and other 
properties of these models are then compared.
The main results on the tree growth model are contained in the sections
describing the growth of the maximal cluster and the cluster containing
the origin. A calculation of the vertex degree correlations in this model is 
the subject of section \ref{correlations}. The final part of
the paper concerns a non-linear generalization of the tree growth
model, which is introduced and numerically simulated in order
to confirm that the cluster growth properties observed in the
tree model are preserved in more complicated models with loops.


\section{Tree Growth Model}
\label{model01}

In each time step a new vertex is introduced.
With probability $\delta$, the new vertex is
connected to another vertex, chosen at random from amongst
the existing vertices.
The vertex remains disconnected with probability $1-\delta$.
In numerical simulations we always start with a single vertex at
time $t = 1$, but we do not expect this initial condition
to affect results at large times.

This model only generates clusters of tree graphs. 
There is only a single tree cluster for the case $\delta = 1$.
These clusters are fragile in the sense that single deletions will always
destroy connectivity~\cite{Barabasi1}.

\subsection{Degree Distribution}

We commence by investigating the distribution of the vertex degrees,
that is, the number of links attached to a given vertex.
Following the notation and methods of Callaway {\it et al.}, we
denote the expected number of vertices of degree $k$ at time $t$
by $d_k(t)$. Since the total number of vertices at time $t$ is
precisely $t$, the probability of attaching a new link to an existing
vertex of degree $k$ is $d_k/t$, leading to the following
evolution equations:
\begin{eqnarray}
\label{eq_01_d_0}
d_0(t+1) &=& d_0(t) - \delta {d_0(t)\over t} + (1-\delta)\\
\label{eq_01_d_1}
d_1(t+1) &=& d_1(t) - \delta {d_1(t)\over t} + 
\delta {d_0(t)\over t} + \delta\\
\label{eq_01_d_k}
d_k(t+1) &=& d_k(t) - \delta {d_k(t)\over t} + 
\delta {d_{k-1}(t)\over t}, \qquad k\ge 2
\end{eqnarray}
Note that the total number of vertices can be written as
$\sum_0^\infty d_k(t) = t$ and that the total expected number of links is
given by $\frac12 \sum_0^\infty kd_k(t) = \delta t$. 
Since both quantities grow linearly in time we search for solutions 
of the form, $d_k(t) = p_k t$, and find:
\begin{eqnarray}
\label{eq_01_p_0}
p_0 &=& {1-\delta \over  1+\delta}\\
\label{eq_01_p_k}
p_k &=& {2\over  1+\delta }\left({\delta\over 1+\delta}\right)^k, 
\qquad k\ge 1
\end{eqnarray}
This distribution decays exponentially in contrast to
random graph models which have a Poisson degree distribution, and 
the scale free models with power law distribution~\cite{Barabasi2}.

\subsection{Static Analog - Branching Process}

Before proceeding to investigate clustering issues we
pause to introduce a non-grown or static analog of this model. 
The static model should have the same vertex degree distribution
as the grown model, but should be constructed to avoid any correlation
between the degree of linked vertices that might arise from the growing
process. 
Furthermore, the analog should preserve the
tree-like character of the model,
so it cannot be one of the classic random graphs of 
Erd\H{o}s and R\'enyi~\cite{RandomGraph}.
An appropriate model is based on an ensemble of Galton and Watson
branching processes~\cite{Feller,A and Ney}.

A branching process may be regarded as a growth process in its
own right, but each vertex is treated identically,
thus avoiding any potential correlation between vertex degrees.
In order to reproduce the vertex degree distribution, 
we choose the probability of $k$ offspring to be proportional to
$p_{k+1}$ in equations (\ref{eq_01_p_0},\ref{eq_01_p_k}), so:
\begin{equation}
\label{eq_static_p_k}
p_k = {1\over  1+\delta }\left({\delta\over 1+\delta}\right)^k, 
\qquad k\ge 0
\end{equation}
This choice gives the correct ratios of vertex degrees at all
higher levels. However, at the first level, where no link is already
present, it is not obvious that the choice correctly weights
the vertices with no children at all. We return to this issue
when we discuss the ensemble of branching processes. 

The properties of the model 
are then a textbook exercise~\cite{Feller},
but for completeness we summarize the main steps.
The main concern is the with the cluster sizes, in particular the
question of percolation. This approach based on branching processes
is identical to the studies of percolation on trees, for
example Bethe lattices, which were
popular in the 1980's \cite{Stauffer} and provided a mean field model
for the percolation transition. 

Percolation occurs in this model when
the extinction probability of the branching process is less than
unity. This extinction probability 
may be calculated using the 
generating function for the probabilities
(\ref{eq_static_p_k}):
\begin{equation}
g(x) = \sum_0^\infty p_k x^k
= {1\over 1+\delta - \delta x}
\label{eq_static_g}
\end{equation}
The extinction probability is 
given by the smallest root, $x_0$, of the equation: 
$g(x) = x$. This root is 1 for all values of $\delta$
so percolation never takes place (though, in the same way as for
one dimensional percolation, $\delta=1$ may be regarded as a critical point).

The technique above can be extended to find the distribution
$n_i$, of finite clusters in this model.
For a single branching process, the generating function, 
$\rho(x) = \sum_1^\infty n^B_i x^i$, 
for the quantities $n^B_i$, which are the probabilities that
the process contains $i$ nodes, is given by the solution to,
$\rho(x) = xg(\rho(x))$, and is found to be:
\begin{equation}
\rho(x) = 
{(1+\delta)\over 2\delta}
-{1\over 2\delta}\sqrt{(1+\delta)^2 - 4\delta x}
\label{eq_static_rho}
\end{equation}
The quantities $n^B_i$ may now be read off, however these are not
the cluster numbers $n_i$, as usually defined. 
The static model is an ensemble of branching processes, 
so $n^B_i$ corresponds to
the number of clusters of size $i$ per process, but $n_i$ is
the number per node. To relate these quantities we compute
the average number of nodes in a branching process as
$\rho\prime(1) = 1/(1-\delta)$. In the limit of a large ensemble
we then find $n_i = (1-\delta)n^B_i$. 
A proper discussion of the ensemble would allow a number of
isolated nodes besides the clusters based on branching processes,
in order to adjust the degree distribution. 
This more careful discussion leads to the same result. 
\begin{eqnarray}
\label{eq_static_n_1}
n_1 &=& {1-\delta\over 1+\delta}\\
\label{eq_static_n_2}
n_2 &=& {\delta(1-\delta)\over(1+\delta)^3}\\
\label{eq_static_n_3}
n_3 &=& {2\delta^2(1-\delta)\over (1+\delta)^5}
\end{eqnarray}
A recursion relation may be obtained for higher order terms.
These results are used for comparison with the tree growth model.

\subsection{Cluster Size Distribution}

\begin{figure}
\centerline{\hbox{\epsfig{figure=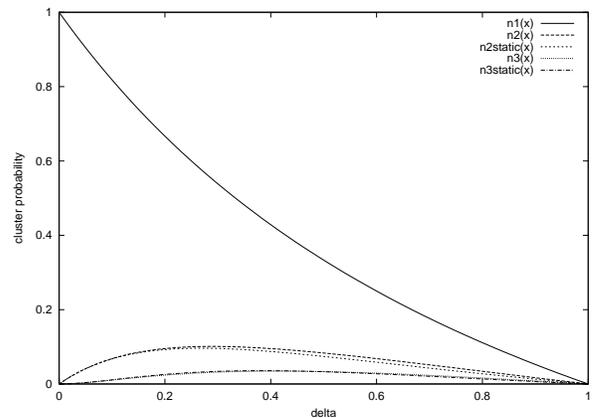,width=8cm}}}
\caption{The fraction of vertices in clusters of
size 1,2,3 ($n_1,n_2,n_3$) according to the formulae
(\ref{eq_01_n_1},\ref{eq_01_n_i}) 
and also the static results 
(\ref{eq_static_n_1},\ref{eq_static_n_2},\ref{eq_static_n_3}). 
These predictions coincide for clusters
of size 1. Simulation results lie on the exact curve but
are not shown in this plot.}
\label{fig_clustersize}
\end{figure}

The expected number of clusters of size $i$, $N_i$,
in the tree growth model obey a
set of evolution equations that can be obtained by noting that
the probability of making a link to a cluster of $i$ vertices is
$iN_i/t$. In contrast to the situation in more complicated models,
these equations are linear, exact and hold for finite $t$.
\begin{eqnarray}
\label{eq_01_N_1}
N_1(t+1) &=& N_1(t) - \delta {N_1(t)\over t} + (1-\delta)\\
\label{eq_01_N_i}
N_i(t+1) &=& N_i(t) - \delta {i N_i(t)\over t} + 
\delta {(i-1) N_{i-1}(t)\over t}
\end{eqnarray}
The expected total number of
clusters $\sum_1^\infty N_i(t)$ 
grows linearly in time and is given by $(1-\delta) t$,
since a new cluster is created whenever a link is not
made in a time step. 
By summing the equations (weighted by $i$), we also find
that the first moment is given by the total number of vertices,
$\sum_1^\infty i N_i(t) = t$. 
These relations also reflect the fact that 
each cluster is a tree graph, so the number of links is 
the number of vertices minus one. 
We search for the cluster size distribution, $n_i$,
of the form, $N_i(t) = n_i t$ and find the following
recursion relations:
\begin{eqnarray}
\label{eq_01_n_1}
n_1 &=& {1-\delta\over 1+\delta}\\
\label{eq_01_n_i}
n_i &=& {(i-1)\delta\over(i\delta+1)}\, n_{i-1} \qquad {i \ge 2}
\end{eqnarray}
Although the first term, $n_1$, is (by design) 
the same as for the static model, later
terms are different. Figure \ref{fig_clustersize}
shows the first few terms of the cluster size distribution for 
both the static model and the growth model.
Notice that while the exact result is similar to the static one
for small delta, it is smaller for larger delta. 

Indeed, the large cluster behavior of the static and the growth
model are completely different. The large cluster behavior of
(\ref{eq_01_n_i}) is power law:
\begin{equation}
n_i 
\stackrel{i \to \infty}{\longrightarrow} 
n_1 \Gamma(2+1/\delta)  i^{-(1+1/\delta)}
\label{eq_01_niscaleing}
\end{equation}
That of the branching model is dominated by exponential decay.

The power law decay of the cluster distribution for the growing
model has been noticed by Dorogovtsev {\it et al.}~\cite{Mendes2} in the
non-percolating phase of non-linear growth models. They have termed
it a self organized critical state because the preferential
attachment to larger clusters which causes the power law
decay occurs automatically. 
For non-grown networks,
exponential decay of the cluster numbers is a common feature 
(as in the static example).
This difference has consequences for the way clusters grow.

\section{Percolation}

Direct numerical simulations of the growing network indicate that
for values of $\delta$ larger than about $1/2$, there is a cluster
of size considerably larger than the others.
This cluster often contains the original
vertex and suggests that there may be a percolating cluster based
on the ``old core'' of vertices that are created early in the
growth. These numerical simulations are in fact misleading, 
but expose anomalous finite size effects that are studied
below.

An analytic approach to percolation does not take the usual
route because
the equations (\ref{eq_01_N_1},\ref{eq_01_N_i}) are exact
and hold for any incipient percolating cluster besides the finite clusters.
Ordinarily, the sum, $\sum_1^\infty i n_i$ only accounts for finite
clusters and the infinite cluster must be added separately.
However, according to the equations, this sum equals $t$ and contains all
the vertices, thereby leaving no room for an infinite cluster.
The generating function approach used in~\cite{Newman01a},
although pleasantly tractable, merely reproduces this information.


Percolation does not occur in this model, except in the trivial limiting
case $\delta  = 1$ where the network just consists of a single tree graph.
This phase diagram resembles that of ordinary one dimensional percolation.
To understand the reasons why percolation does not take place, yet
large clusters do appear at finite size, it is
helpful to study the numerical data for the maximum sized cluster.
This will then lead us to an investigation of the cluster containing
the original site.

\subsection{Numerical Study}

\begin{figure}
\centerline{\hbox{\epsfig{figure=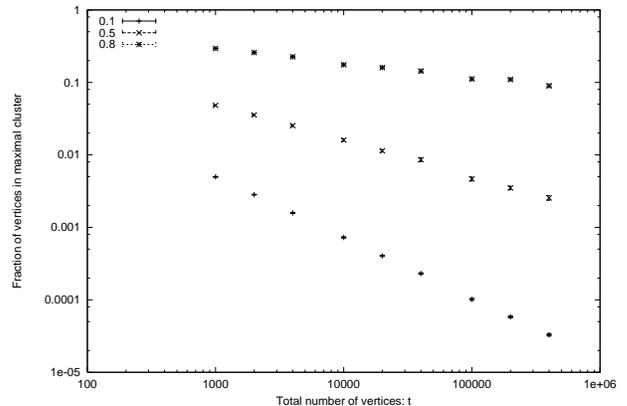,width=8cm}}}
\caption{Scaling of largest cluster size fraction against $t$
for $\delta = 0.1,0.5,0.8$}
\label{fig_Nscaletree}
\end{figure}

On closer inspection
of the numerical data it is found that the fraction of
sites contained in the largest cluster suffers from an
anomalously slow finite size
effect, becoming smaller as the growth
process is continued to larger times. 
For example, at $\delta = 0.8$, the fraction drops from about
0.29 at $t=10^3$, to 0.17 for a network 10 times larger.
In figure \ref{fig_Nscaletree}
we show the 
largest cluster fraction against $\log (t)$ for various $\delta$.
The straight lines clearly indicate a power law dependence.
The exponent can be determined by fitting, or by noticing that
another plot of the same quantity ($\log$) against $\delta$ 
displays linear dependence. In any event, the lack of any
transition is clear. The fit suggests
the form:
\begin{equation}
{\rm Fraction\ of\ vertices\ in\ largest\ cluster} \sim t^{\delta-1}
\label{eq_01_scaling}
\end{equation}
This form of scaling behavior can be deduced from the original growth model.
Consider a large isolated cluster, $N_{\bar i} = 1$.
By treating its size, ${\bar i}$, as a continuous variable,
we find that it grows according to the probability
that a link will attach the new vertex to this cluster:
\begin{equation}
{\bar i}(t+1) = {\bar i}(t) + \delta {{\bar i}\over t} 
\label{eq_01_imax}
\end{equation}
There is no solution linear in $t$, but a form 
${\bar i} \sim t^\delta$ solves the equation in the large time
limit. The fraction of sites in this largest cluster,
${\bar i}/t$, therefore follows the scaling behavior observed
numerically in (\ref{eq_01_scaling}). As the system grows
very large, the relative size of even the largest cluster
decreases and it is apparent that the tree growth model never 
experiences true percolation.

In most static models with percolation, for example random graph
models, the finite size scaling of
the maximum cluster size is given by $\log(t)$. This is
related to the usual exponential decay of the cluster 
size distribution, and the power law behavior we see
here follows from the distinctive decay (\ref{eq_01_niscaleing})
in growth models.


\subsection{Cluster Containing the Initial Point}

The overall distribution of the sizes of randomly chosen
clusters (\ref{eq_01_n_1},\ref{eq_01_n_i})
does not give any hint of the presence of
the large cluster seen in the numerical work above.
It is hard to investigate the maximal cluster analytically,
but if we rely on the observation that the maximal cluster is likely to
be based on one of the oldest vertices, we
may approach the problem from a different perspective.
The distribution of the size of clusters
that contain the original point is amenable to analytic
methods and does shed some light on the presence of a large cluster.
The possibility of studying this quantity is of
course only available in grown networks that have distinguished
vertices.

As before, we start by writing evolution equations,
this time for the probability $P_i(t)$ that a
distinguished cluster has size $i$
at time $t$ ($1\le i \le t$). 
\begin{eqnarray}
\label{eq_01_P_1}
P_1(t+1) &=& P_1(t) - \delta {P_1(t)\over t}\\
\label{eq_01_P_i}
P_i(t+1) &=& P_i(t) - \delta {i P_i(t)\over t} 
+  \delta  {(i-1) P_{i-1}(t)\over t}\\
\label{eq_01_P_t}
P_{t+1}(t+1) &=& \delta P_t(t)
\end{eqnarray}
These equations are very similar to the ones for the overall 
cluster size distribution $N_i(t)$ in (\ref{eq_01_N_1},\ref{eq_01_N_i}),
however, the difference in the first equation prevents any solution
$P_i(t) \propto t$. 
The equations actually hold for {\it any} distinguished
cluster, with the initial condition determining
which cluster is selected.
Simplest is to choose the cluster
distinguished as containing the original point, in which
case $P_1(1) = 1$. Other possibilities, for example
the cluster containing the second point would be determined
by the values at $t=2$, $P_1(2) = 1-\delta$ and $P_2(2) = \delta$.
This in fact leads to the same distribution as for the first point,
but a difference is obtained for the third point which is specified by:
$P_1(3) = 1-\delta$, $P_2(3) = \delta(1-\delta)$ and $P_3(3) = \delta^2$.
In the following, we shall only consider the cluster containing the original
point.

The sum $\sum_1^t P_k(t)$ is preserved by these equations,
and can be set to 1, as expected for a probability, by the initial condition.
The average size of the distinguished cluster, 
${\bar k}(t) = \sum_1^t kP_k(t)$, obeys
${\bar k}(t+1) = (1+\delta/t){\bar k}(t)$. So at large
times we expect that ${\bar k}(t) \sim t^\delta$.
This is essentially the same argument as in (\ref{eq_01_imax})
of the last section and indeed the evolution equation has the
same intuitive origin. In this form the prefactor can be determined
from the initial condition.
Evolution equations for all the higher moments of the 
distribution will be considered below.

For large $t$ and $k$, the continuum version of the evolution
equation becomes:
\begin{equation}
t{\partial P\over \partial t}
=
-\delta\, {\partial (kP)\over \partial k}
\end{equation}
which has a scaling solution,
\begin{equation}
P(t,k) 
=
t^{-\delta} f (kt^{-\delta})
\end{equation}
where $f(u)$ is any function.

\begin{figure}
\centerline{\hbox{\epsfig{figure=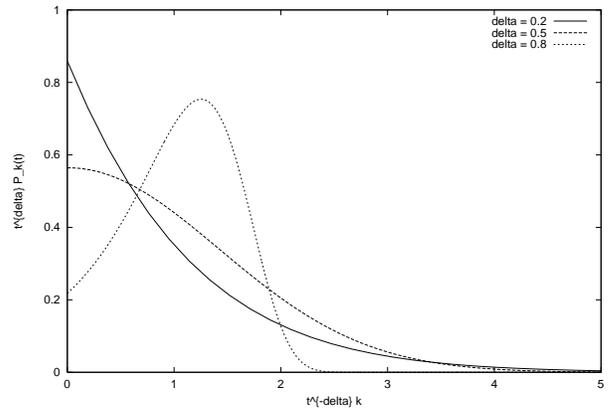,width=8cm}}}
\caption{Size distribution of the cluster containing the origin
with a scaling plot of $t^{\delta}P_k(t)$ against $t^{-\delta}k$.
For a variety of values of $\delta$. The lines are obtained by
numerically solving the equations 
(\ref{eq_01_P_1},\ref{eq_01_P_i},\ref{eq_01_P_t}) up to $t=10^4$.}
\label{fig_scaling}
\end{figure}

This result is confirmed, and the form of the scaling function
$f(u)$ determined, by numerically solving the difference
equations (\ref{eq_01_P_1},\ref{eq_01_P_i},\ref{eq_01_P_t}) 
and plotting them appropriately
as shown in figure \ref{fig_scaling}.
No change in the form of the function is visible as $t$
is increased beyond about 2000.
A scaling relation of this form is interesting because it
is found for all values of $\delta$ not just those in the vicinity
of the critical point at $\delta = 1$.

The scaled cluster distribution shows a clear
change in form around $\delta = 1/2$. Although the mean of the
distribution varies smoothly with
$\delta$, and is close to 1 on the scaled plot
(corresponding to ${\bar k}(t) = t^\delta$ before scaling), the
mode moves away from zero (cluster size, $k=1$, before rescaling)
as $\delta$ becomes greater than about $1/2$. 
Eventually, as $\delta \to 1$ the scaling function becomes 
progressively more peaked around $u =1$.
This provides an argument for the likely presence of
a large maximal cluster for  $\delta \gtrsim 1/2$. 

The form of the scaling
function is not easy to determine analytically. 
Only in the limit of large or small $u$, can $f(u)$ be determined
using the solutions for $P_1(t)$ and $P_t(t)$ obtained from 
(\ref{eq_01_P_1}) and (\ref{eq_01_P_t}).

For comparison with simulations it is better to compare
the moments of the distribution rather than the full form.
The moments, defined as,
\begin{equation}
S_n(t) = \sum_1^t k^n P_k(t)
\end{equation}
obey simple equations obtained from weighted  sums of 
(\ref{eq_01_P_1},\ref{eq_01_P_i},\ref{eq_01_P_t}).
\begin{eqnarray}
S_0(t+1) &=& S_0(t)\\
S_1(t+1) &=& (1 + {\delta\over t})S_1(t)\\
S_2(t+1) &=& (1 + {2\delta\over t})S_2(t) + {\delta\over t}S_1(t)
\end{eqnarray}
And similar equations for higher order moments.
By forming suitable linear combinations, these equations
can be solved in terms of the following function:
\begin{eqnarray}
R(z,t)& =& \prod_{i=1}^{t-1} (1 + z/i) = {\Gamma(z+t)\over \Gamma(t) \Gamma(z + 1)}\\
&\stackrel{t \to \infty}{\longrightarrow} & {t^z\over \Gamma(z+1)}
\end{eqnarray}
For example:
\begin{eqnarray}
S_0(t) &=& R(0,t) = 1\\
S_1(t) &=& R(\delta,t)\\
S_2(t) &=& 2R(2\delta,t) - R(\delta,t) 
\end{eqnarray}
The general case is not difficult to work out, and it is also
possible to treat clusters containing other than the original
point.
As $t$ becomes large, $R(n\delta,t) \sim t^{n\delta}$, so the
leading term dominates and $S_n(t) \to n!R(n\delta,t)$.
However, for finite $t$, the sub-leading terms are large
in the region $\delta \lesssim 1/\log(t)$ and must be kept in
numerical work.

\begin{figure}
\centerline{\hbox{\epsfig{figure=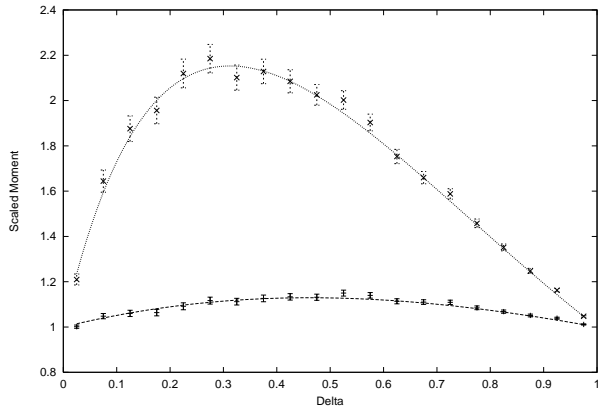,width=8cm}}}
\caption{First two moments of the size of cluster containing the origin.
The upper curve is for the second moment, and the lower curve is for
the mean.
Each moment is scaled according to $t^{-n\delta}S_n(t)$ and
based on simulations of 4000 samples of networks containing $10^4$ vertices.
The lines are the theoretical results including the sub-leading
finite size terms.}
\label{fig_moments}
\end{figure}

In figure \ref{fig_moments}
we show comparisons of these formulae against simulation
results for the mean and the second moment.
Bearing in mind the scaling behavior, we plot each moment divided
by a power of $t^\delta$. These first moments show excellent
agreement.


\section{Vertex Degree Correlations}
\label{correlations}

To conclude our study of the tree growth model 
we follow the same argument used by Calloway {\it et al.}
to determine the correlations between the vertex degree at
each end of a randomly chosen link. The number of edges that
join vertices of degrees $j$ and $k$ is denoted $E_{jk}$.
This matrix is symmetric. 
For links that join vertices of the same degree, 
$E_{kk}$ is defined to be {\it twice} the number of such links.
In this case {\it exact} evolution equations can be derived by
treating the vertices with a single link specially:
\begin{eqnarray}
\label{eq_01_E_11}
E_{11}(t+1) &=& E_{11}(t) 
+ 2 \delta {d_0\over t} 
- 2 \delta p_1 {E_{11}\over d_1}\\
\label{eq_01_E_1k}
E_{1k}(t+1) &=& E_{1k}(t) 
+ \delta {d_{k-1}\over t} 
- \delta \left( p_1 {E_{1k}\over d_1} + p_k {E_{1k}\over d_k} \right)\\
\label{eq_01_E_jk}
E_{jk}(t+1) &=& E_{jk}(t) 
+ \delta \left( p_{k-1} {E_{jk-1}\over d_{k-1}} + p_{j-1} {E_{j-1k}\over d_{j-1}} \right)\nonumber \\
&-& \delta \left( p_j {E_{jk}\over d_j} + p_k {E_{jk}\over d_k} \right)
\end{eqnarray}
Where the $d_k(t)$ and $p_k$ are the vertex degree numbers and their
probabilities as determined earlier in section \ref{model01}.

The total expected number of links
is given by $\frac12\sum_{jk} E_{jk}(t)$
and the evolution equations show that it is given by $\delta t$
as anticipated. We therfore write the probabilities as
$E_{jk}(t) = 2\delta t e_{jk}$, and derive the following equations.
\begin{eqnarray}
\label{eq_01_e_11}
(1 + 2\delta) e_{11} &=&  p_0\\
\label{eq_01_e_1k}
(1 + 2\delta) e_{1k} &=&  {p_{k-1}\over 2} 
+ \delta\, e_{1k-1}\\
\label{eq_01_e_jk}
(1 + 2\delta) e_{jk} &=&  
\delta \left( e_{jk-1} + e_{j-1k} \right)
\end{eqnarray}

By appropriately multiplying these equations and adding, we can 
find the following relations between the moments:
\begin{eqnarray}
M_0 &=& \sum_{jk} e_{jk}  = 1\\
M_1 &=& \sum_{jk} j e_{jk} = 1 + \delta + \frac12 \sum_{j=0} jp_j \\
M_2 &=& \sum_{jk} jk e_{jk} = 1 + 2\delta M_1 + \sum_{j=0} jp_j
\end{eqnarray}
Using the results of section \ref{model01} on the vertex degrees,
we find the average degree $\sum_0^\infty kp_k = {\bar k} = 2\delta$.  
The sum above includes vertices with no links, and 
the average degree on the end of a randomly chosen link is,
$\mu = \sum k^2 p_k / \sum k p_k = 1 + 2\delta$. 

The covariance between vertex degrees at each end of a randomly chosen
link is defined as:
\begin{eqnarray}
C &=& \sum_{jk} (j-\mu)(k-\mu)e_{jk}\\
  &=& M_2 - 2\mu M_1 + \mu^2 M_0
\end{eqnarray}
Combining these results we find that $C$ vanishes identically and that there
is no correlation between the degrees at the end of randomly chosen 
links in this model.
This result is supported by simulations.

In view of this result, 
it is slightly surprising that the analog static model which
was specifically designed to avoid these correlations, is not
identical to the tree graph model. There is still a distinction
as was apparent from the cluster numbers.



\section{Two Link Growth Model}
\label{model02}

The preceding study of cluster growth in the 
tree growth network has been reasonably tractable, fundamentally
due to the tree property of the network. The question
arises as to which features
are preserved in more general models. 

The most obvious 
difference in more complicated models is the presence of a 
percolating phase.
The tree growth model has no
percolating phase except the trivial one at $\delta = 1$. The
physical reason for this deficiency is not directly the tree
nature of the network.
The cause should rather be sought in the growth
itself. There is no mechanism to attach existing
clusters to each other. A mechanism of this type was responsible for
the percolating properties in the model of Calloway {\it et al.},
and clearly introduces non-linearities into the model, for example
in the equation for the generating function of cluster sizes.

A natural extension of our tree growth model is a model in
which at each time step a new vertex is created and then,
with probability $\delta$ connected by {\it two} links to 
the existing vertices. Each new link is assigned a random
terminating vertex amongst the existing vertices.
The networks grown by this model are not necessarily tree-like
and loops can form. The static analog is therefore a random graph
rather than a branching process.
Again, in numerical work, we use an initial condition consisting
of a single vertex.
A further generalization that we have considered is a model
in which there is are fixed probabilities for single and
double link connections. This leads to a more complicated
phase diagram, but not to any significantly new observations.

Below, we present the basic properties of the two link model
closely following the methods of Calloway {\it et al.}. We then
perform simulations to
study the cluster properties and compare them with what was
found in the tree model.

\subsection{Two Link Growth Model - Degree Distribution}

The equations leading to the degree distribution 
that are obtained by the same means as for the tree growth model.
\begin{eqnarray}
\label{eq_02_d_0}
d_0(t+1) &=& d_0(t) - 2\delta {d_0(t)\over t} + (1-\delta)\\
\label{eq_02_d_1}
d_1(t+1) &=& d_1(t) - 2\delta {d_1(t)\over t} + 
2\delta {d_0(t)\over t}\\
\label{eq_02_d_2}
d_2(t+1) &=& d_2(t) - 2\delta {d_2(t)\over t} + 
2\delta {d_1(t)\over t} + \delta\\
\label{eq_02_d_k}
d_k(t+1) &=& d_k(t) - 2\delta {d_k(t)\over t} + 
2\delta {d_{k-1}(t)\over t}
\end{eqnarray}
Note that the total number of vertices, t, can be written as
$\sum_0^\infty d_k(t)$ and that the expected number of links, 
$2\delta t$, is
given by $\frac12 \sum_0^\infty kd_k(t)$. 
Searching for solutions of the form, $d_k(t) = p_k t$, we find:
\begin{eqnarray}
p_0 &=& {1-\delta\over 1+2\delta}\\
p_1 &=& 2\delta(1-\delta)\over (1+2\delta)^2\\
p_k &=& (1+8\delta){2^{k-2}\delta^{k-1}\over (1+2\delta)^{k+1}} 
\qquad ({\rm for}\ k\ge 2)
\label{eq_02_p_k}
\end{eqnarray}
Again this distribution decays exponentially after the first couple
of terms.

\subsection{Two Link Growth Model - Cluster Size Distribution}

The cluster sizes $N_i$ in the two link growth model obey a
set of evolution equations which are now approximate and 
only valid for finite clusters at large $t$ since processes in 
which both links end in the same cluster are ignored. This is the
same approximation that is made in the Calloway {\it et al.} analysis.
\begin{eqnarray}
\label{eq_02_N_1}
N_1(t+1) &=& N_1(t) - 2\delta {N_1(t)\over t} + (1-\delta)\\
\label{eq_02_N_2}
N_2(t+1) &=& N_2(t) - 2\delta {2 N_2(t)\over t}\\
\label{eq_02_N_i}
N_i(t+1) &=& N_i(t) - 2\delta {i N_i(t)\over t}\nonumber\\  
&+&\delta \sum_{j=1}^{i-2}{j N_j(t)\over t}{(i-j-1) N_{i-j-1}(t)\over t}
\end{eqnarray}
Solutions of the form $N_i(t) = n_i t$ are considered and a recursion 
relation obtained:
\begin{eqnarray}
n_1 &=& {\delta\over 1+2\delta}\\
n_2 &=& 0\\
n_i &=& {\delta\over 1+2i\delta}\sum_{j=1}^{i-2}j n_j (i-j-1) n_{i-j-1}
\label{eq_02_n_i}
\end{eqnarray}
Analysis of these cluster numbers is best carried out using the
generating function for the cluster sizes, $g(x) = \sum_1^\infty i n_i x^i$,
which obeys a non-linear equation:
\begin{equation}
g' = {1\over2\delta} 
\left({1-\delta - g/x +  \delta g^2\over 1 - xg} \right)
\label{eq_02_g}
\end{equation}

\subsection{Two Link Growth Model - Percolation}

\begin{figure}
\centerline{\hbox{\epsfig{figure=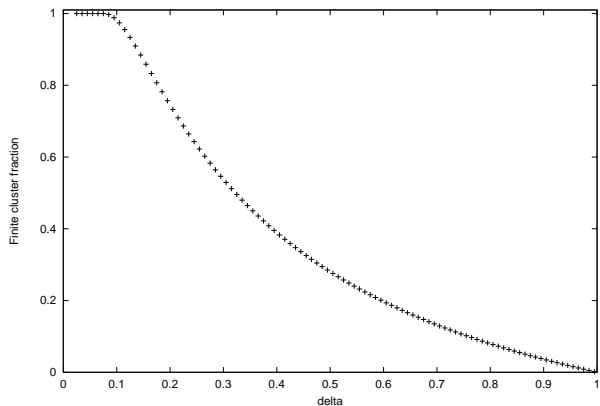,width=8cm}}}
\caption{
The fraction of the vertices contained within finite clusters.
Obtained by numerical integration of the differential
equation \ref{eq_02_g} using a step size of $10^{-6}$.}
\label{fig_twolink}
\end{figure}

We compute $g(1)$ by numerically integrating the equation
(\ref{eq_02_g}) starting from an initial condition 
$(g(\epsilon) = n_1 \epsilon)$.
Figure \ref{fig_twolink} shows the results,
and we recall that $g(1)$ is the expected fraction of vertices
contained in the
finite clusters, so when it differs from 1, percolation 
occurs. 

The model percolates for most of the range of $\delta$, 
but for a range of small  $\delta$ there is no percolation.
It is possible to obtain the critical value $\delta_c$
by  studying $g'(1)$.
In the percolating region $g(1) < 1$, so it is simple to
take the $x \to 1$ limit of the right hand side of equation
(\ref{eq_02_g}) to obtain,
\begin{equation}
g'(1) = {1\over 2\delta} \left( 1 - \delta (1 + g(1))\right)
\label{eq_02_gprime1}
\end{equation}
In the case  $\delta < \delta_c$,  $g(1) = 1$, and this limit must
be taken more carefully with the help of L'Hopital's rule. 
The resulting quadratic equation can be solved to give:
\begin{equation}
g'(1) = {1 - 4\delta 
\pm \sqrt{1- 16 \delta + 16\delta^2}
\over   4\delta }
\label{eq_02_gprime2}
\end{equation}
We omit regions where the root is not real, and further
require that it be positive. Finally, recognizing that
$g(1) \to 1$ as $\delta \to 0$ since in this limit all
clusters have size one, we are able to pick the negative sign
as being the only correct branch.

In summary: $\delta_c = 1/2 - \sqrt{3}/4 \approx 0.06699$. With $g'(1)$
taking different values on each side:
\begin{equation}
g'(1) =
\biggl\lbrace
\begin{array}{ll}
{1 - 4\delta 
- \sqrt{1- 16 \delta + 16\delta^2}
\over   4\delta } 
&\qquad {\rm for}\   \delta < \delta_c\\
{1\over 2\delta} \left( 1 - \delta (1 + g(1))\right) 
&\qquad {\rm for}\   \delta > \delta_c
\end{array}
\label{eq_02_gprime3}
\end{equation}
%

The critical behavior we have described is very similar to
that observed in the model studied by  Calloway {\it et al.}. 
By performing a similar investigation near the critical 
point, we find the same signals of an infinite order transition with
$1-g'(1) \sim e^{\alpha/\sqrt{\delta - \delta_c}}$.



\section{Cluster Growth in the Two Link Growth Model}

In this section we describe the results of numerical simulations
to find how large clusters grow in this model. We track both
the maximal cluster and the cluster containing the original vertex.

\begin{figure}
\centerline{\hbox{\epsfig{figure=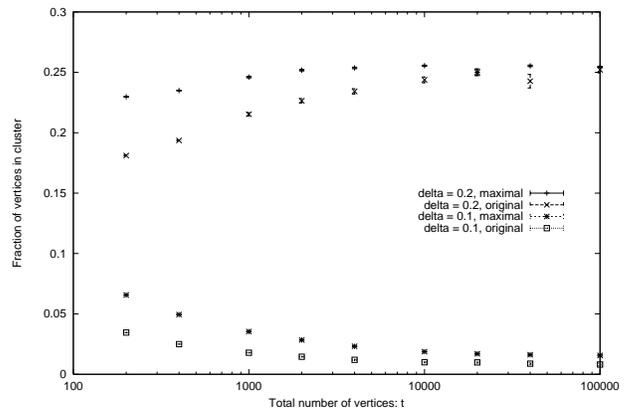,width=8cm}}}
\caption{Maximal and original cluster fractions in the
two link model for two values of $\delta$, ($0.1,0.2$) within the percolating region.
In each case the maximal curve is above the original one.
Averages are taken over a large
number of samples ranging from $10^2$ for the largest networks
to $10^5$ at the smallest.}
\label{fig_twolinkaboveperc}
\end{figure}

In the region above the percolation threshold the maximal cluster
naturally grows with $t$. It is interesting to see how 
finite size affects influence this and how the
cluster containing the original point grows. This is shown in 
figure \ref{fig_twolinkaboveperc} which indicates that there
is a region where the original cluster is smaller
than the maximal one, but as the size of the network increases,
this cluster approaches the size of the maximal one. This 
result supports the intuition that the ``old core'' of vertices
act as a seed for the percolating cluster. Indeed, the probability
that the maximal cluster contains the original vertex appears
to grow to 1 for any $\delta$ in the percolating phase.
Unfortunately the statistics for this analysis are not good
for the sizes we have considered and this result should only
be taken as suggestive.

The finite size effects are most apparent for $\delta = 0.1$
which is quite close to the critical point. 
In this case the fraction of sites in the either maximal or
original cluster decrease with $t$ in a way reminiscent of
the behavior in the tree growth model.
Estimates of a
correlation size can be made on the basis of logarithmic plots
which show a clear change in slope as the network size exceeds 
the correlation size at that value of $\delta$.

\begin{figure}
\centerline{\hbox{\epsfig{figure=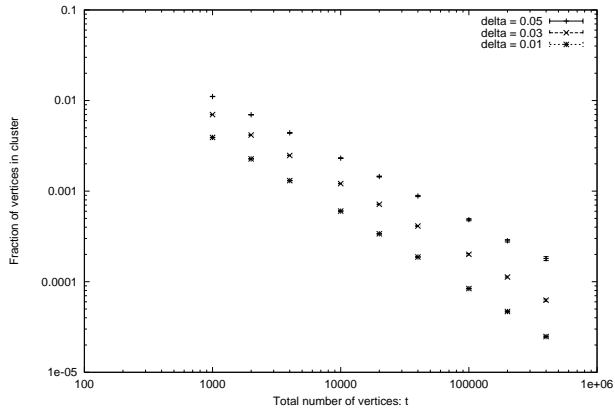,width=8cm}}}
\caption{Scaling of maximal cluster in the
two link model below the percolation threshold.
The original cluster follows similar curves.
For $\delta = 0.01,0.03,0.05$. Averages  are taken over a large
number of samples ranging from $50$ for the largest networks
to $10^4$ for the smallest.}
\label{fig_twolinkbelowperc}
\end{figure}

It is the situation below the percolation threshold that 
holds more interest for comparison with the tree growth model.
In figure \ref{fig_twolinkbelowperc} we show evidence that
the maximal cluster scales with a power law
decay in this region. The original cluster behaves in the
same way.
This is exactly as in the tree growth model,
and as emphasized before, quite distinct from the $\log(t)$
behavior in random graph models.

Having demonstrated that scaling occurs in the same way as
in the tree growth model,
we postpone any further study of the exponent of the growth.
This is because of the difficulty of getting far
from the critical point in this particular model.



\section{Conclusion}
\label{conclusion}

The study of grown networks was originally motivated by real networks
which are by nature finite. We have shown in a simple 
tree growth model, that although an
infinite size system does not display percolation, finite
systems of sizes that may have relevance to observations,
often contain large clusters. 
These clusters grow with a power law dependence on the system
size and provide another manifestation of the critical nature
of the whole phase.
The power law growth can be analysed carefully in this model,
especially by studying clusters with distinguished points,
but the pattern of power law growth appears to be general
as found in numerical simulations in a non-linear model.

\section{Acknowledgments}
I would like to thank the PDC group in Southampton University for the use of 
computer facilities.


\end{document}